\documentclass[aps,prb,twocolumn,superscriptaddress,nofootinbib,longbibliography,floatfix]{revtex4-2}

\usepackage{amsmath,amssymb,amsfonts,bm,mathtools}
\usepackage{graphicx}
\usepackage{xcolor}
\usepackage[colorlinks=true,linkcolor=blue,citecolor=blue,urlcolor=blue]{hyperref}
\usepackage{microtype}
\usepackage{booktabs}
\usepackage{float}
\graphicspath{{figures/generated/}{figures/}}

\newcommand{\rr}{\mathbf r}
\newcommand{\RR}{\mathbf R}

\newcommand{\CS}{\mathrm{CS}}
\newcommand{\PF}{\mathrm{F}}
\newcommand{\CF}{\mathrm{CF}}
\newcommand{\cm}{\mathrm{CM}}
\newcommand{\sgn}{\operatorname{sgn}}
\newcommand{\cH}{\mathcal H}
\newcommand{\cF}{\mathcal F}
\newcommand{\cB}{\mathcal B}
\newcommand{\dd}{\mathrm d}
\newcommand{\ii}{\mathrm i}
\newcommand{\ee}{\mathrm e}
\newcommand{\Tr}{\operatorname{Tr}}
\newcommand{\llB}{\ell_B}

\begin{document}

\title{Neural Flux Attachment: From Bose Condensates to Chiral Topological Matter}
\hypersetup{
  pdftitle={ChernFormer: Neural Flux Attachment from Bose Condensates to Chiral Topological Matter},
  pdfauthor={Author names to be inserted},
  pdfsubject={Neural quantum states, Chern-Simons transmutation, Bose condensation, chiral topological matter}
}

\author{Rudik Badalyan}
\affiliation{Department of Physics, University of Massachusetts, Amherst, Massachusetts 01003, USA}

\author{Khachatur G. Nazaryan} 
\affiliation{Department of Physics, Massachusetts Institute of Technology, Cambridge, MA 02139, USA}

\author{Tigran A. Sedrakyan}
\affiliation{Department of Physics, University of Massachusetts, Amherst, Massachusetts 01003, USA}
\affiliation{A. Alikhanyan National Science Laboratory, Yerevan 0036, Armenia}

\begin{abstract}
Can one neural wave function describe both a Bose condensate and a chiral topological liquid? We introduce ChernFormer, which combines a fermionic transformer with a fixed Chern-Simons phase that attaches one statistical vortex to every particle pair. Each factor changes sign under exchange, so their product is exactly bosonic. The fixed phase changes statistics but not probability, making every bosonic learning problem equivalent to a fermionic one with the same approximation error and overlap. With enough capacity, ChernFormer can approximate any normalizable bosonic wave function on the plane at fixed particle number. A finite, smooth network still vanishes when particles meet, yet this contact hole can shrink while the wave function and condensate fraction approach those of a nodeless condensate. 
Following the needle-in-a-haystack target-reconstruction benchmark introduced in \cite{NazaryanGaggioliTengFu2025}, we test ChernFormer on the Kalmeyer--Laughlin ground state and its first two chiral edge states.
The overlap curves stay close to unity through their largest sampled sizes, while independent amplitude and phase maps at $N=20$ for all three states recover both local Laughlin vortices and the collective edge vortex. By contrast, a continuous, nonzero product of identical particle-wise factors  misses these elementary edge sectors. ChernFormer therefore provides one variational language for conventional bosonic order and chiral topological matter.
\end{abstract}

\maketitle

\section{Introduction}
\label{sec:intro}

A variational calculation can find only states that its trial wave function is able to represent. Variational Monte Carlo makes this idea practical by optimizing a parametrized many-body wave function and using Monte Carlo samples to evaluate the energy and other observables \cite{Sorella2005}. Neural-network variational Monte Carlo enlarges the search space by replacing a short analytic ansatz with a trainable complex function of all particle coordinates or occupation numbers \cite{CarleoTroyer2017,Carleo2019}. Neural wave functions now reach high accuracy for molecules and continuum electrons \cite{Pfau2020, Hermann2020, vonGlehn2023, GaggioliFu2026, GaggioliGrahamFu2026}, frustrated magnets \cite{Viteritti2023}, interacting lattice bosons \cite{Denis2025}, strongly correlated systems \cite{Geier2025, NazaryanFu2026, ZGFu2026, AGGFu2026, LPOCLFu2026}, and systems with spin or other internal degrees of freedom \cite{Avdoshkin2026}. This progress sharpens a basic physics question: does the chosen architecture contain every phase that the Hamiltonian may favor, or does its built-in structure exclude some phases before optimization begins? Recent calculations for the Hofstadter--Bose--Hubbard model, where bosons move in an effective magnetic field, found a pronounced loss of neural-state accuracy as the magnetic flux increased across several optimizers and network families \cite{Ledinauskas2025}. The choice of architecture can therefore be as important as the choice of optimizer.

Bosonic matter is an unusually demanding test because one exchange symmetry supports very different forms of order. In a Bose--Einstein condensate, a macroscopic fraction of the particles occupies one coherent orbital, and strong short-range correlations can coexist with this long-range coherence \cite{PenroseOnsager1956,ReattoChester1967,Reatto1969}. A crystal selects a periodic density pattern, while a supersolid combines density order with a coherent quantum phase \cite{Chester1970,Bottcher2019,Tanzi2019}. Bosonic fractional quantum Hall liquids and Kalmeyer--Laughlin chiral spin liquids are different again. They contain correlation holes, a complex many-body phase, topological order--a global form of quantum organization not described by a local order parameter---and chiral edge modes that propagate in one direction \cite{Laughlin1983,KalmeyerLaughlin1987,Wen1989,Wen1990}. Microscopic calculations have identified the Kalmeyer--Laughlin phase in frustrated magnets, constructed its edge wave functions, and recently resolved the chiral edge of a bosonic fractional Chern insulator at filling one-half \cite{HeShengChen2014,Herwerth2015,YangChenDong2026}. Exact bosonic symmetry is common to all of these states, but it does not determine which form of order is present.

Neural wave functions have already captured important parts of this landscape. Nonlocal neural states can represent chiral topological wave functions, self-attention can learn fractional quantum Hall correlations, and recent attention-based calculations can discover topological degeneracy directly from energy minimization \cite{Glasser2018,Teng2025,Abouelkomsan2026}. The remaining issue is architectural. A network may reproduce the density and local pair correlations of a target while still missing its global phase sector. Here, a phase winding is the integer number of full turns made by the complex wave-function phase along a closed path through many-particle configurations. For chiral matter, this global information is essential because edge excitations are distinguished by collective phase winding, not only by local correlation holes.

\begin{figure}[!t]
\centering
\includegraphics[width=\columnwidth]{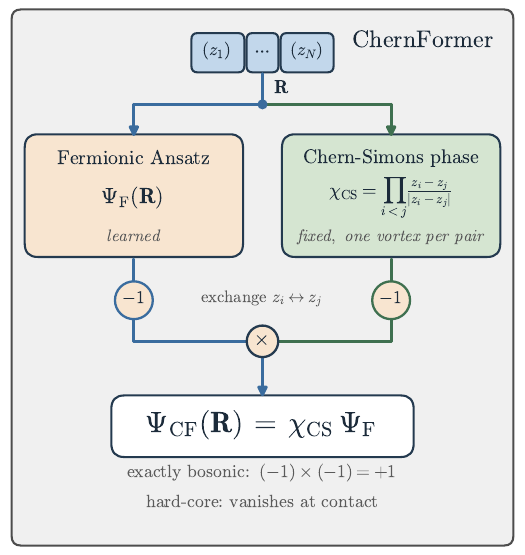}
\caption{Neural flux attachment in ChernFormer.  A trainable Fermionic Backbone is antisymmetric under particle exchange, while the fixed pairwise Chern--Simons factor is also antisymmetric and has unit magnitude.  Their product is therefore exactly bosonic and has the same probability density as the fermionic backend.  For a smooth finite backend, antisymmetry also enforces a contact zero, giving a hard-core bosonic architecture while leaving the long-distance order and remaining many-body phase trainable.}
\label{fig:architectureoverview}
\end{figure}

ChernFormer addresses this problem by separating exchange statistics from the correlations that must be learned. In two dimensions, attaching a phase vortex to each particle pair can change the exchange statistics while leaving the probability density unchanged. This idea grew from the theory of two-dimensional particle statistics and underlies Chern--Simons descriptions of the fractional quantum Hall effect and composite particles \cite{LeinaasMyrheim1977,Wilczek1982,ZhangHanssonKivelson1989,Jain1989}. Related constructions have been developed for interacting bosons and quantum magnets \cite{Sedrakyan2012,Sedrakyan2015,Wang2018,Wang2022,SedrakyanGlazmanKamenev2015,SedrakyanGlazmanKamenev2014,MaitiSedrakyan2019,WeiSedrakyan2023,SedrakyanMoessnerKamenev2020}, and for synthetic gauge fields and cold-atom systems \cite{ValentiRojas2020,ValentiRojas2023,Kamal2024}. ChernFormer implements flux attachment directly in the wave function. It multiplies a fixed pair phase by a complex fermionic ansatz. The pair phase and the fermionic determinant each change sign when two particles are exchanged, so their product is exactly bosonic. Because the fixed factor has magnitude one, it changes only the phase and leaves the probability density unchanged. The neural determinant is then free to learn the physical amplitude and all remaining many-body phase structure. Figure~\ref{fig:architectureoverview} gives the construction at a glance.  The antisymmetric fermionic backbone and the antisymmetric pairwise statistics layer contribute one exchange sign each, so the two signs cancel exactly.  Because the fixed layer has unit magnitude, it fixes only exchange statistics: the trainable backend still carries the probability density, correlations, and residual many-body phase.
Furthermore, one of us (Nazaryan) and collaborators introduced a "needle-in-a-haystack" benchmark in which a neural network learns a chosen complex many-body wave function from its probability density and probability current \cite{NazaryanGaggioliTengFu2025}. Recent works further shows that one parameter-conditioned self-attention model can represent a family of interacting-electron ground states across Hamiltonian parameters \cite{NazaryanFu2026, ZGFu2026}.

This construction raises three concrete questions. For a specified Fermionic Backbone family, what is the exact class of bosonic wave functions that can be represented? Does the zero inherited from fermionic antisymmetry prevent an accurate description of a smooth condensate whose amplitude is finite when two particles meet? And can the same architecture learn the collective winding of a chiral edge, rather than only the vortices attached to individual particle pairs?

We answer all three questions. (i) Away from particle collisions, multiplication by the Chern--Simons phase is a unitary, or norm-preserving, change of statistics. The best bosonic approximation error is therefore exactly the best fermionic approximation error for the target obtained after removing the fixed pair phase. With an arbitrarily expressive complex Fermionic Backbone, every normalizable bosonic state on the plane or disk can be approached in integrated norm at fixed particle number. (ii) Every smooth finite ChernFormer has an exact contact zero. This is naturally matched to hard-core and Laughlin-like states. A weakly interacting condensate with finite contact amplitude is instead approached through a sequence with an increasingly narrow correlation hole. The many-body norm and condensate fraction converge, and in two dimensions the energy can also converge for smooth, nonsingular Hamiltonians. The short-distance contact rule therefore does not select the long-distance order; we call this contact--order separation.

(iii) The Kalmeyer--Laughlin edge tower provides a sharp global test. For a droplet of $N$ bosons, its center-of-mass edge descendants are labeled by a non-negative integer $p$, which counts the added edge angular momentum and phase winding. ChernFormer can carry every integer $p$. An equivariant product made from $N$ identical marked-particle factors can carry the same winding only in multiples of $N$, as long as the wave function remains continuous and nonzero on the diagnostic region. We train ChernFormer on the ground state and the first two edge states, $p=0,1,2$. The global-overlap curves remain close to unity through $N=20$ for $p=0,1,2$. The phase-resolved test  at $N=20$ shows that the  amplitude and phase maps reproduce the local Laughlin vortices and, for $p=1,2$, the additional collective edge vortex with one and two units of winding. The edge results therefore test not only numerical accuracy but also access to the correct many-body phase sector.

In Section~\ref{sec:architecture}, we introduce ChernFormer and argue its exact representability for many-body bosonic wavefunctions. In Section~\ref{sec:condensates}, we explain its applicability to condensates, discussing contact order separation and ordered states. Sections~\ref{sec:KL} and \ref{sec:training} develop the Kalmeyer--Laughlin edge benchmark and present the training results. In Section~\ref{sec:discussion}, we summarize the physics and the remaining numerical tests. The appendices contain the formal derivations and implementation diagnostics.

\section{ChernFormer: neural flux attachment}
\label{sec:architecture}

\subsection{The fermionic neural backbone}

We consider $N$ identical particles moving in a two-dimensional region.  The position of particle $i$ is $\rr_i=(x_i,y_i)$, and we combine its two Cartesian coordinates into the complex number $z_i=x_i+\ii y_i$.  The full many-body configuration is $\RR=(z_1,\ldots,z_N)$.  The labels $1,\ldots,N$ are only bookkeeping devices: a physical bosonic wave function must be unchanged when any two labels are exchanged.

Fermionic Backbone first turns the coordinates and pair separations into a feature vector for each particle.  Self-attention then allows the feature of every particle to use information from every other particle.  The same update rule is used for all labels.  Therefore, relabeling the input particles simply relabels the output features in the same way.  This property is called \emph{permutation equivariance}; it is the key symmetry of the neural layers.

To write the Fermionic Backbone output compactly, let $\mathcal E(\RR)$ be a symmetric prefactor that controls the large-distance envelope and may include known symmetric pair correlations.  Let $M$ be the number of determinant channels, let $c_\alpha$ be their complex weights, and let $\phi^{(\alpha)}_{\mu j}(\RR)$ be the complex entry in orbital row $\mu$ and particle column $j$ of channel $\alpha$.  A generic complex Fermionic Backbone has the schematic form
\begin{equation}
 \Psi_{\PF}(\RR)=\mathcal E(\RR)
 \sum_{\alpha=1}^{M}c_\alpha
 \det\!\left[\phi^{(\alpha)}_{\mu j}(\RR)\right].
 \label{eq:psiformer}
\end{equation}
Each generalized orbital can depend on the complete configuration $\RR$, rather than only on the coordinate $z_j$ in its own column.  The orbitals can therefore adapt to the instantaneous many-particle environment.  In traditional many-body language, this environment dependence is called backflow.  Here it means simply that the orbitals are genuine many-body functions rather than fixed one-particle shapes.

When two particles are exchanged, permutation equivariance exchanges the corresponding two columns of every orbital matrix.  A determinant changes sign when two columns are exchanged, while the symmetric prefactor $\mathcal E$ does not.  Eq.~\eqref{eq:psiformer} is therefore exactly antisymmetric.  The determinant enforces fermionic exchange statistics analytically; the attention network learns the amplitude, correlations, nodes, and complex phase.  Several determinants are not needed for the symmetry itself, but they give the state more freedom to describe complicated nodal and phase structures.

\subsection{Attaching one statistical vortex to every pair}

For a pair of particles, write the relative displacement as $z_i-z_j=r_{ij}\exp(\ii\theta_{ij})$.  Here $r_{ij}=|z_i-z_j|$ is the pair distance and $\theta_{ij}$ is its polar angle.  The ratio $(z_i-z_j)/|z_i-z_j|=\exp(\ii\theta_{ij})$ is therefore a pure phase: it records only the direction of the pair and has unit magnitude.

ChernFormer multiplies the Fermionic Backbone by one such phase for every pair,
\begin{equation}
 \begin{gathered}
 \chi_{\CS}(\RR)=\prod_{i<j}\frac{z_i-z_j}{|z_i-z_j|},\\[-2pt]
 \Psi_{\CF}(\RR)=\chi_{\CS}(\RR)\Psi_{\PF}(\RR).
 \end{gathered}
 \label{eq:main_ansatz}
\end{equation}
Eq.~\eqref{eq:main_ansatz} is the central definition of the architecture.  The product of all pair differences,
$\Delta(\RR)=\prod_{i<j}(z_i-z_j)$, is often called the Vandermonde factor. With this shorthand, $\chi_{\CS}=\Delta/|\Delta|$.

The exchange symmetry follows directly.  Exchanging two particles rotates their relative coordinate by half a turn, so the corresponding pair phase changes by $\exp(\ii\pi)=-1$.  The remaining pair factors are only relabeled.  Thus $\chi_{\CS}$ changes sign under an odd exchange.  The Fermionic Backbone determinant changes sign as well, and the two signs cancel.  The ChernFormer wave function is therefore exactly bosonic.

The same phase has a simple flux-attachment interpretation.  If one particle makes a complete counterclockwise circuit around another, $\theta_{ij}$ increases by $2\pi$.  The factor $\chi_{\CS}$ supplies one unit of local phase winding, or one statistical vortex, to that pair.  This changes the exchange statistics without changing the probability density because
\begin{equation}
 |\Psi_{\CF}(\RR)|^2=|\Psi_{\PF}(\RR)|^2.
 \label{eq:same_density}
\end{equation}
The pair angle is undefined when two coordinates coincide, so Eq.~\eqref{eq:main_ansatz} is first defined on configurations with no particle collisions.  The contact limit is not ignored: the antisymmetric backend fixes it and is analyzed explicitly in Sec.~\ref{sec:condensates}.

 Particle coordinates and pair separations enter permutation-equivariant attention layers.  Those layers produce complex, configuration-dependent orbital matrices.  Determinants turn the matrices into an antisymmetric fermionic amplitude.  Finally, the fixed Chern--Simons phase in Eq.~\eqref{eq:main_ansatz} converts that amplitude into an exactly symmetric bosonic wave function.  The analytic layer supplies only the required statistical vortex; all trainable density structure and all additional phase structure remain in Fermionic Backbone.

\subsection{The exact representability}

Let $\Psi_B(\RR)$ be a target bosonic wave function.  Removing the known statistical phase gives the inverse-transmuted target
\begin{equation}
 \Psi_F^{\rm target}(\RR)=\chi_{\CS}^{*}(\RR)\Psi_B(\RR),
 \label{eq:inverse_target}
\end{equation}
where the star denotes complex conjugation.  Under particle exchange, $\Psi_B$ remains unchanged while $\chi_{\CS}^{*}$ changes sign.  Their product $\Psi_F^{\rm target}$ is therefore antisymmetric.  A bosonic target is represented by a chosen ChernFormer exactly when the corresponding Fermionic Backbone can represent the fermionic function in Eq.~\eqref{eq:inverse_target}.  This is the complete classification for any fixed Fermionic Backbone architecture.

The same relation makes the learning problem especially transparent.  Let $\Psi_\theta$ denote the trainable Fermionic Backbone, with $\theta$ collecting all neural-network parameters.  The usual many-body $L^2$ norm, $\|\Psi\|_2$, is 
$
\left[
\int d^2\mathbf r_1\cdots d^2\mathbf r_N,
\bigl|\Psi(\mathbf r_1,\ldots,\mathbf r_N)\bigr|^2
\right]^{1/2}$. Since $|\chi_{\CS}|=1$ away from collisions,
\begin{equation}
 \|\Psi_B-\chi_{\CS}\Psi_\theta\|_2
 =\|\chi_{\CS}^{*}\Psi_B-\Psi_\theta\|_2.
 \label{eq:isometry}
\end{equation}
Equation~\eqref{eq:isometry} is the key approximation result.  Training ChernFormer on a bosonic target is exactly equivalent, at the level of integrated wave-function error, to training Fermionic Backbone on its inverse-transmuted fermionic partner.  Normalized overlaps are also unchanged, and therefore so is the fidelity, which is the squared normalized overlap.  The fixed statistics layer neither improves nor degrades the best possible global fit.

Under the standard universal-approximation assumptions, a complex, fully antisymmetric generalized-determinant Fermionic Backbone with unrestricted capacity can approximate any square-integrable fermionic wave function.  Equation~\eqref{eq:isometry} then implies that its ChernFormer image can approximate any normalizable symmetric bosonic wave function on the plane or disk.  In this precise global sense, condensates, crystals, supersolids, current-carrying states, and chiral topological liquids all belong to the same limiting family.  The appendices \ref{app:map},\ref{app:determinant} give the unitary map and a constructive determinant representation on compact regions away from collisions.

This completeness statement concerns the available function class, not the cost of learning it.  A finite network can still be too small, can use an unsuitable envelope, or can omit a required spatial symmetry.  A real-valued backend cannot reproduce a general chiral phase.  On a torus or sphere, the planar pair phase in Eq.~\eqref{eq:main_ansatz} must be replaced by a phase compatible with the boundary geometry.  Most importantly, a smooth finite determinant has a specific contact zero.  The next section explains why this local constraint does not prevent high-accuracy learning of long-range bosonic order.

A second bosonic ansatz is obtained by squaring the fermionic output,
\begin{equation*}
 \Psi_{\rm sq}(\RR)=\bigl[\Psi_{\PF}(\RR)\bigr]^2.
\end{equation*}
The exchange sign is squared away, but the exact function class is smaller.  A bosonic target is included only if it has a single-valued antisymmetric square root that the chosen Fermionic Backbone can represent.  Thus, on every closed loop where the state remains nonzero and all labels return to their starting points, the phase winding is even; isolated zeros generated by the square also have even multiplicity. ChernFormer does not impose a square-root condition: its unit-magnitude phase gives the linear, norm-preserving map in Eq.~\eqref{eq:isometry}.  It therefore retains the full bosonic $L^2$ expressive power of the fermionic backend.

\section{Condensates}
\label{sec:condensates}

\subsection{Can ChernFormer learn a condensate?}

An ideal condensate with fixed particle number places every boson in the same normalized orbital $\phi(\rr)$ within a two-dimensional region $D$.
\begin{equation}
 \int_D |\phi(\rr)|^2\dd^2r=1,
 \qquad
 \Phi_0(\RR)=\prod_{i=1}^{N}\phi(\rr_i).
 \label{eq:hartree}
\end{equation}
The first relation normalizes the orbital.  The product in the second relation is unchanged by particle exchange, so $\Phi_0$ is bosonic.

The one-body density matrix $\gamma^{(1)}(\rr,\rr')$ measures how much phase coherence remains between $\rr'$ and $\rr$ after the other $N-1$ particles have been integrated out.  Its eigenvalues give the occupations of coherent one-particle orbitals.  We call the largest occupation $n_0$; the ratio $n_0/N$ is the condensate fraction and equals one for Eq.~\eqref{eq:hartree}.

At first sight, the fermionic backbone appears to rule out this nodeless state.  An antisymmetric wave function must vanish when two particle coordinates coincide, and the unit-magnitude Chern--Simons phase cannot remove that zero. 
If $P_{ij}\mathbf R$ denotes the configuration obtained by exchanging particles $i$ and $j$. Fermionic antisymmetry gives $\Psi_{\rm PF}(P_{ij}\mathbf R)=-\Psi_{\rm PF}(\mathbf R)$. At coincidence, the exchanged configuration is identical to the original one, so antisymmetry requires $\Psi_{\rm PF}=0$. For nearby particles, a regular finite Fermionic Backbone, such as PsiFormer, is locally Lipschitz: its change under a small displacement of the full configuration is bounded by a finite local slope. The configuration-space distance between $\mathbf R$ and $P_{ij}\mathbf R$ is $\sqrt{2}|\mathbf r_i-\mathbf r_j|$. Combining this distance with antisymmetry gives
$
|\Psi_{\rm PF}(\mathbf R)|
\leq C|\mathbf r_i-\mathbf r_j|.
$
  A finite ChernFormer therefore obeys
\begin{equation}
 \begin{aligned}
 \Psi_{\PF}(\ldots,\rr_i=\rr_j,\ldots)&=0,\\
 |\Psi_{\CF}(\RR)|&\leq C|\rr_i-\rr_j|
 \quad (\rr_i\to\rr_j),
 \end{aligned}
 \label{eq:linearcontact}
\end{equation}
where $C$ is a finite local slope.  Although the pair phase itself is undefined at an exact collision, the full product has a unique continuous extension with value zero.  The normalized pair correlation $g^{(2)}(r)$ compares the probability density for a pair at separation $r$ with the value expected from the one-particle densities.  Equation~\eqref{eq:linearcontact} therefore gives $g^{(2)}(0)=0$: every smooth finite ChernFormer has an exact contact hole.

This contact rule is a natural match to hard-core and Laughlin-like bosons.  A hard-core state already assigns zero amplitude to configurations in which two particles occupy the same point, so the determinant enforces the physical boundary condition rather than an artificial one.  The same statement is explicit for one Laughlin pair factor.  Removing one unit of statistical phase gives
\begin{equation}
 \frac{(z_i-z_j)^*}{|z_i-z_j|}(z_i-z_j)^2
 =(z_i-z_j)|z_i-z_j|.
 \label{eq:laughlininversepair}
\end{equation}
The right-hand side is antisymmetric, continuous, and zero at contact.  It is therefore a regular fermionic target for Fermionic Backbone.  A weakly interacting condensate is different because its exact amplitude remains finite when two particles meet.  No single smooth finite ChernFormer can reproduce that contact value pointwise.  The correct limiting description is instead the shrinking-hole sequence introduced next.

This local rule does not prevent almost perfect condensation, because condensation concerns coherence over macroscopic distances.  To separate the two scales, let $f_{\delta,\epsilon}(r)$ be a pair factor that is zero for $r\leq\delta$, one for $r\geq\epsilon$, and smooth in between.  The smaller length $\delta$ is the core radius of the hole, while $\epsilon$ is the distance over which the wave function returns to the condensate.  Starting from $\Phi_0$, define
\begin{equation}
 \Phi_{\delta,\epsilon}(\RR)=
 \mathcal N_{\delta,\epsilon}\Phi_0(\RR)
 \prod_{i<j}f_{\delta,\epsilon}(|\rr_i-\rr_j|),
 \label{eq:condensatesequence}
\end{equation}
where $\mathcal N_{\delta,\epsilon}$ restores unit norm.  Every state in this family has the required contact zero, but it is identical to the target condensate whenever all pair separations exceed $\epsilon$.

\begin{figure}[!t]
\centering
\includegraphics[width=\columnwidth]{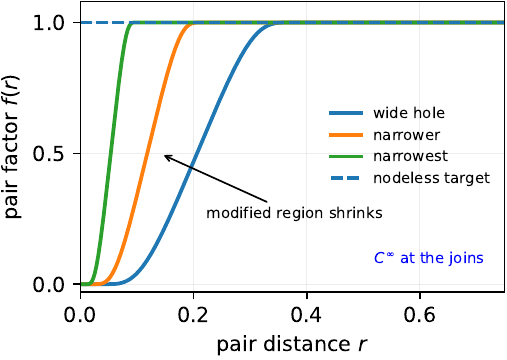}
\caption{A shrinking contact hole using $C^\infty$ (the set of smooth, infinitely differentiable functions) pair factors.  Each curve is zero in a small core and equals the nodeless target beyond the healing distance $\epsilon$, with smooth joins.  The piecewise logarithmic profile used for the bound is an $H^1$ proof device and can be mollified with arbitrarily small changes in $L^2$ and gradient norm; a smooth neural backend need not reproduce a cusp.}
\label{fig:contactorder}
\end{figure}

The nonsmooth behavior is not a feature of the neural-network ansatz. The piecewise logarithmic function introduced in Appendix~\ref{app:condensate} is used only as a convenient mathematical tool to prove the convergence properties. Since smooth functions can approximate any $H^1$ function arbitrarily well, the two joining points of this regulator can be smoothly rounded without changing the important results. In particular, the shrinking region, the $L^2$ convergence, and the kinetic-energy scaling remain unchanged. Therefore, a smooth Fermionic Backbone learns a smooth contact regulator, as illustrated in Fig.~\ref{fig:contactorder}, rather than a function with sharp corners or discontinuous derivatives.

Figure~\ref{fig:contactorder} shows the construction for one pair.  In two dimensions, the area within distance $\epsilon$ of a particle scales as $\epsilon^2$.  At fixed $N$, the probability that any pair enters this shrinking region therefore vanishes.  The normalization approaches one, the many-body wave function converges in integrated norm, and the one-body density matrix converges with it.  The central condensate result is
\begin{equation}
 \|\Phi_{\delta,\epsilon}-\Phi_0\|_2\longrightarrow0,
 \qquad
 \frac{n_0}{N}\longrightarrow1.
 \label{eq:l2condensate}
\end{equation}
The $L^2$ norm integrates the wave-function difference over all particle coordinates.  The convergence is not pointwise: every finite $\Phi_{\delta,\epsilon}$ still vanishes at contact.  Instead, the configurations on which it differs from $\Phi_0$ carry vanishing weight.  Nearly complete condensation and exact suppression of zero-separation pairs can therefore coexist.

Energy convergence is a stronger requirement because kinetic energy depends on derivatives.  Two dimensions are favorable: for the logarithmic healing profile given in Appendix~\ref{app:condensate}, the gradient cost of one hole is
\begin{equation}
 \int_{\delta<r<\epsilon}\dd^2r\,
 |\nabla f_{\delta,\epsilon}(r)|^2
 =\frac{2\pi}{\ln(\epsilon/\delta)}
 \longrightarrow0.
 \label{eq:logenergy}
\end{equation}
Here $\epsilon\to0$ while $\delta$ shrinks faster, so $\epsilon/\delta\to\infty$.  For smooth confining potentials and nonsingular interactions, and provided the neural network also learns the required derivatives, the variational energy can approach the condensate energy.  An idealized zero-range interaction probes the contact region directly and needs a separate analysis.

Equations~\eqref{eq:linearcontact}, \eqref{eq:l2condensate}, and \eqref{eq:logenergy} capture the main physics: the statistical vortex fixes the exact behavior at zero separation but not the order at long distances.  We call this \emph{contact--order separation}.  Local pair probes remain sensitive to the hole, while interference, the momentum distribution, and the one-body density matrix can show an almost ideal condensate.  Appendix~\ref{app:condensate} gives the full norm, reduced-density-matrix, and energy estimates.

\subsection{Which ordered states are covered?}

The result of the previous subsection is that ChernFormer fixes the
short-distance contact rule but does not choose the long-distance order.
The Chern--Simons factor and the antisymmetric Fermionic Backbone backend make
the wave function bosonic; for a regular finite backend they also impose
a zero when two particles meet. Away from these collision points, the
complex backend learns the physical amplitude and the remaining
many-body phase. Thus, at fixed particle number on the plane or disk, an
unrestricted backend can approximate any normalizable bosonic state in
the sense discussed in Sec.~II, with the finite-network contact caveat
discussed above.

This class includes simple and fragmented condensates. A simple
condensate has one macroscopic eigenvalue of the one-body density matrix
$\gamma^{(1)}$, while a fragmented condensate has several. A finite
smooth ChernFormer cannot reproduce the nonzero contact amplitude of an
ideal soft-core condensate pointwise, but the shrinking-hole construction
lets its condensate fraction approach unity. Hard-core bosons and
Laughlin-like liquids are more direct targets, because their physical
wave functions already vanish at contact.

Other forms of order are encoded in the same way. Density order is
carried mainly by the magnitude of the wave function, so crystals,
stripes, droplets, phase separation, vortex lattices, and supersolids are
included. A supersolid is represented when density order coexists with
macroscopic eigenvalues of $\gamma^{(1)}$. Currents, persistent-flow
states, and vortices are encoded in the many-body phase and can be tested
through currents, phase winding, or the response to a phase twist.

Short-range Jastrow correlations are compatible with long-range
coherence \cite{ReattoChester1967,Reatto1969}, and crystalline order can
coexist with condensation in a supersolid
\cite{Chester1970,Bottcher2019,Tanzi2019}. For lattice models, the
planar construction applies directly to hard-core bosons on distinct
occupied sites. Soft-core bosons require a statistics-transmutation
factor written in occupation-number space, because no planar pair angle
is defined when several particles occupy the same site.

ChernFormer is therefore especially useful for states that combine local
correlation holes with nonlocal phase structure. The Kalmeyer--Laughlin
tests below ask whether the backend can learn this collective chiral
phase, edge winding, and topological sector after exchange statistics
have been fixed analytically.


\section{Kalmeyer--Laughlin edge states as an architecture test}
\label{sec:KL}

\subsection{A collective edge excitation}

The Kalmeyer--Laughlin state is the simplest chiral spin liquid and, in the continuum, is the bosonic Laughlin state at filling one half.  On a disk it is
\begin{equation}
 \Psi_{\rm KL}(\RR)=
 \Delta(\RR)^2
 \exp\left[-\sum_i\frac{|z_i|^2}{4\llB^2}\right].
 \label{eq:KL}
\end{equation}
The Vandermonde factor $\Delta=\prod_{i<j}(z_i-z_j)$ was introduced in Sec.~\ref{sec:architecture}.  Its square is symmetric and vanishes quadratically when two particles meet, giving the Laughlin correlation hole.  Its phase makes two full turns when one particle circles another.  The magnetic length $\ell_B$ sets the size of the Gaussian-confined droplet.  On a lattice, $\Psi_{\rm KL}$ denotes the projected chiral-spin-liquid amplitude with its gauge factors \cite{KalmeyerLaughlin1987,Wen1989,HeShengChen2014,Herwerth2015}.

A clean edge excitation changes only the collective coordinate of the droplet:
\begin{equation}
 \begin{aligned}
 \Psi_p(\RR)&=Z^p\Psi_{\rm KL}(\RR),
 &p&=0,1,2,\ldots,\\
 Z&=\sum_{i=1}^{N}z_i.
 \end{aligned}
 \label{eq:edgefamily}
\end{equation}
The usual center of mass is $Z/N$, and $p$ counts the added center-of-mass angular momentum.  The state $p=0$ is the ground state; $p=1$ and $p=2$ are the first two edge states studied below.  Since $Z$ is symmetric, every $\Psi_p$ remains bosonic.

The factor $Z^p$ acts on the droplet as a whole and leaves every relative coordinate $z_i-z_j$ unchanged.  All members therefore have the same local zeros, correlation hole, and bulk topological order.  They differ at the boundary: $Z^p$ adds a collective zero of order $p$ at $Z=0$, adds $p$ units of angular momentum, and shifts probability weight outward through the factor $|Z|^{2p}$.  In the edge theory this is a chiral density wave, not a bulk quasiparticle, and it is one simple member of the level-$p$ chiral-boson manifold \cite{Wen1990,Herwerth2015,YangChenDong2026}.

Removing the known Chern--Simons phase gives the fermionic target seen by Fermionic Backbone,
\begin{equation}
 \chi_{\CS}^{*}(\RR)\Psi_p(\RR)
 =Z^p\Delta|\Delta|
 \exp\left[-\sum_i\frac{|z_i|^2}{4\llB^2}\right].
 \label{eq:KLpreimage}
\end{equation}
Because $|\Delta|$ is symmetric and $\Delta$ is antisymmetric, this target is fermionic.  The collective factor $Z^p$ passes through the statistics layer unchanged.  Thus one ChernFormer architecture contains the ground state and every non-negative integer member of this center-of-mass edge tower.

If $p=2m$,
\begin{equation*}
 \Psi_{2m}(\RR)=
 \left[
 Z^m\Delta
 \exp\left(-\sum_i\frac{|z_i|^2}{8\ell_B^2}\right)
 \right]^2,
\end{equation*}
so the squared-Fermionic-Backbone ansatz contains the even-$p$ states.  Odd $p$ has odd center-of-mass winding, so any square root is multivalued around $Z=0$ and no exact continuous representation exists.  ChernFormer instead is general and contains every integer $p$.

\subsection{Center-of-mass winding distinguishes the architectures}

The edge quantum number can be read from a phase loop.  Begin with a collision-free configuration $z_i^{(0)}$ whose center of mass is at the origin, $\sum_i z_i^{(0)}=0$, and translate every particle through the same circle,
\begin{equation}
 z_i(t)=z_i^{(0)}+a\ee^{\ii t},
 \qquad 0\leq t\leq2\pi,
 \label{eq:CMloop}
\end{equation}
where $a>0$ is the loop radius.  The droplet moves without rotating or changing shape because all pair separations remain fixed.  We call the number of full $2\pi$ turns made by the continuously unwrapped wave-function phase the center-of-mass winding, $w_{\cm}$.  It is an integer as long as the wave function does not vanish on the loop.

Along this motion, $Z(t)=Na\ee^{\ii t}$, whereas the Laughlin pair factor and the Chern--Simons phase remain constant.  The central result is
\begin{equation}
 w_{\cm}[\Psi_p]
 =w_{\cm}[\chi_{\CS}^{*}\Psi_p]
 =p.
 \label{eq:CMwinding}
\end{equation}
The statistics layer is therefore transparent to rigid center-of-mass motion: PsiFormer carries the complete collective winding, and ChernFormer can access every non-negative integer sector in this edge tower.

An equivariant product uses a different symmetry mechanism,
\begin{equation}
 \Psi_{\rm EP}(\RR)=\prod_{i=1}^{N}
 h\bigl(z_i;\{z_j\}_{j\neq i}\bigr).
 \label{eq:EPmain}
\end{equation}
The same complex function $h$ is used for each marked particle and treats the remaining coordinates as an unordered set, so the product is bosonic.  On the rigid loop, all $N$ factors follow equivalent closed phase paths.  If $h$ is continuous, single valued, and nonzero on the loop and on the deformations that exchange the marked particle, each factor has the same integer winding $m$.  Their phases add:
\begin{equation}
 w_{\cm}[\Psi_{\rm EP}]=Nm,
 \qquad m=0,\pm1,\pm2,\ldots .
 \label{eq:EPwindingmain}
\end{equation}
The product ansatz can therefore carry only $0,\pm N,\pm2N,\ldots$ units of center-of-mass winding.  Appendix~\ref{app:KL} gives the continuity argument.

For every system shown here, $N\geq8$, so the physical $p=1$ and $p=2$ states belong to ChernFormer sectors that a continuous nonzero equivariant product cannot enter.  This is a difference in representable phase sectors, not merely in training quality.  The product can change winding only by developing a zero on the diagnostic loop, where the phase becomes undefined.

Local observables do not reveal this distinction: all $\Psi_p$ have the same pair zeros and local Laughlin winding.  A model can therefore reproduce density and pair correlations while missing the collective edge sector.  A high overlap is also not a complete certificate because a phase slip may occupy a region of very small probability.  
A direct center-of-mass winding test would evaluate the unwrapped phase
and the minimum wave-function magnitude along the rigid translation loop
in Eq.~\ref{eq:EPmain}. In Sec.~\ref{sec:training} we use a slightly different but complementary
phase-resolved diagnostic: we fix $N-1$ particles and scan the remaining
coordinate. This slice is not the pure center-of-mass loop because the
relative coordinates change. Nevertheless, it exposes the same collective
edge factor, since $Z=z+\sum_{j=2}^N z_j$ on the slice. The collective
zero at $Z=0$ therefore appears as a vortex of multiplicity $p$, while
the factors $(z-z_j)^2$ display the local Laughlin vortices.

\section{Learning the Kalmeyer--Laughlin edge tower}
\label{sec:training}

\subsection{Many-body accuracy from the wave-function overlap}

\begin{figure}[!t]
\centering
\includegraphics[width=0.98\columnwidth]{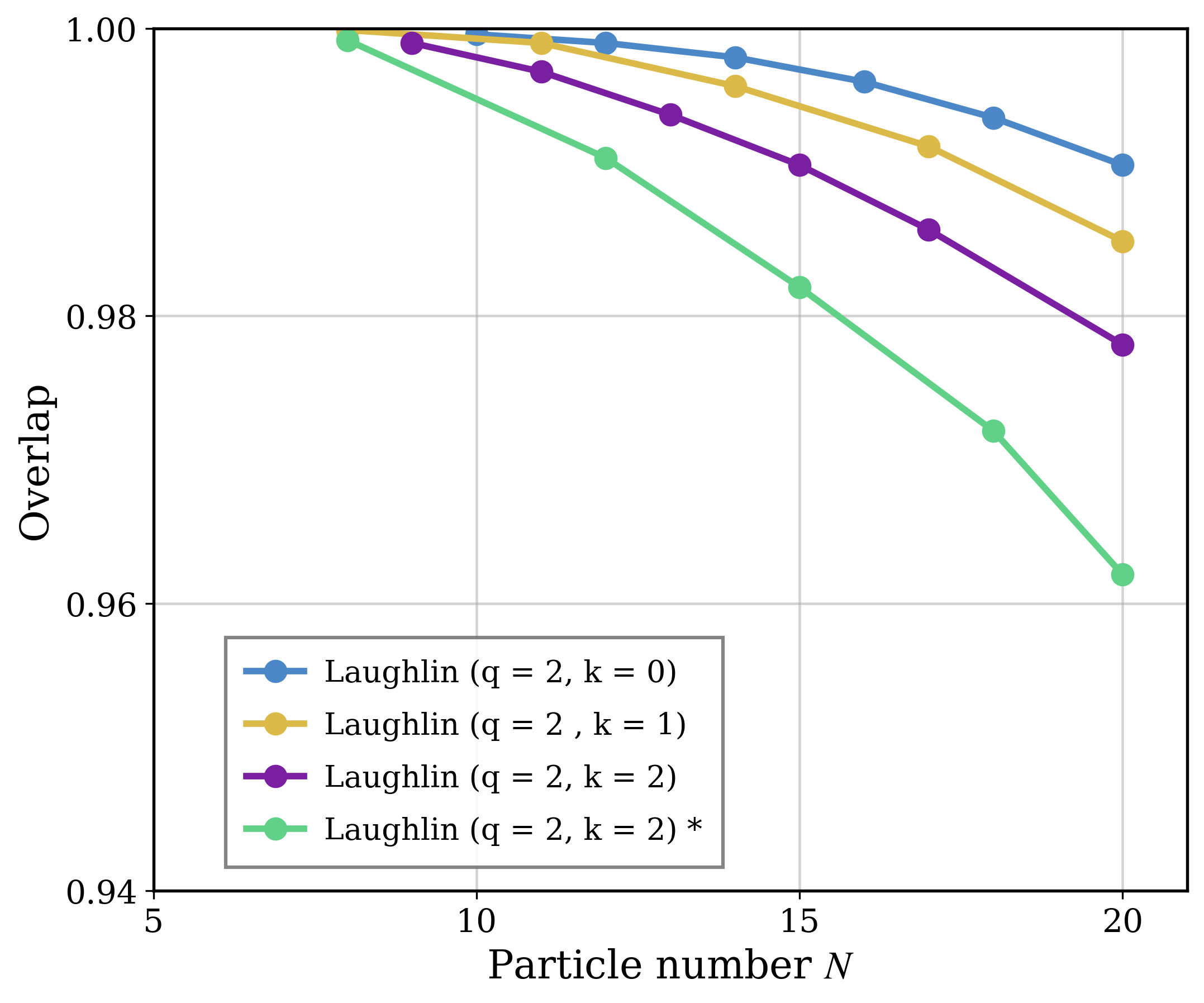}
\caption{Normalized overlap between the target Kalmeyer--Laughlin state and ChernFormer.  Here $q=2$ and $k=p$.  The overlap is approximately $0.990$ for $p=0$ at $N=20$.
Figure~\ref{fig:tomography} gives the corresponding $N=20$ amplitude and phase results for $p=1,2$. The green starred curve is the squared-PsiFormer ansatz $\Psi_{\rm sq}=[\Psi_{\PF}]^2$ for $p=2$ and lies below ChernFormer at the common sizes.}
\label{fig:overlap}
\end{figure}

\begin{figure*}[t]
\centering
\includegraphics[width=\textwidth]{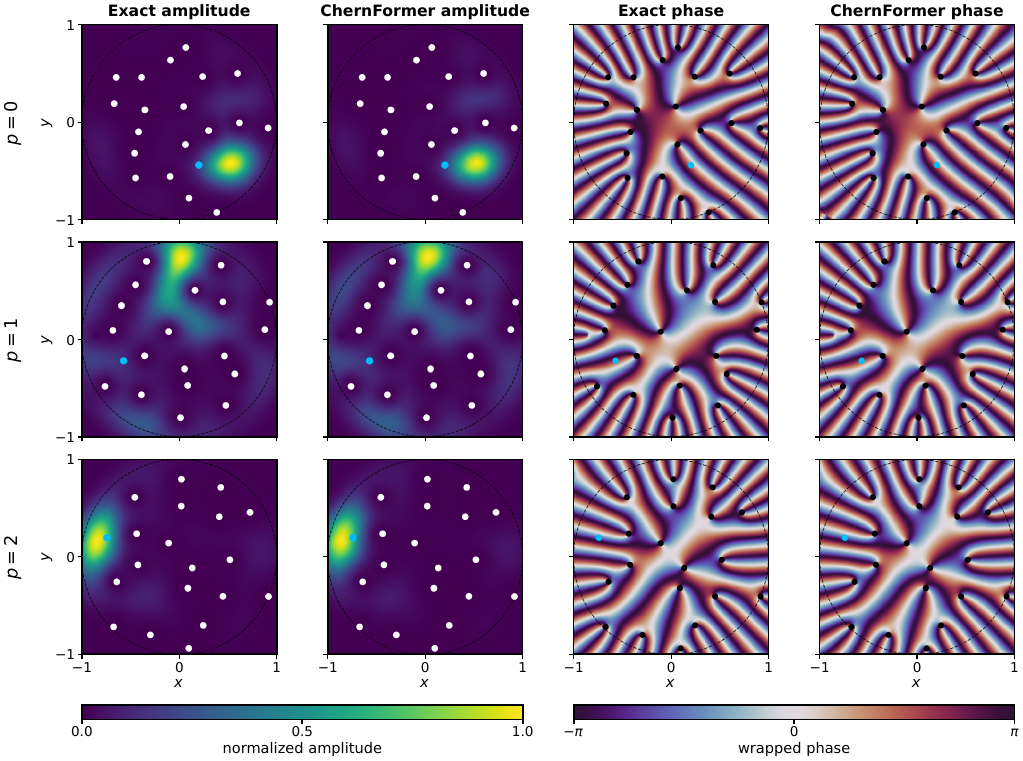}
\caption{Amplitude and phase maps for the $N=20$ Kalmeyer--Laughlin edge tower.    Rows show the three edge indices $p=0$, $1$, and $2$. Columns compare exact and ChernFormer amplitudes and phases.  Fixed particles generate local charge-two Laughlin vortices, while the $p=1$ and $p=2$ rows add a collective zero with one and two units of winding.  Amplitudes are normalized separately in each panel.  The phase is wrapped from $-\pi$ to $\pi$.}
\label{fig:tomography}
\end{figure*}

The states with $p=0$, $1$, and $2$ share the same local Laughlin correlations but carry different collective edge windings.  The overlap curves in Fig.~\ref{fig:overlap} test the full many-body state up to $N=20$ for $p=0$, $1$, and $2$, while Fig.~\ref{fig:tomography} tests all three states at $N=20$ through their amplitude and phase.

For training objective we used needle-in-haystack benchmark introduced in \cite{NazaryanGaggioliTengFu2025}. We shared the training and architectural details in the Appendix \ref{app:TrainingDetails}.

For a target $\Psi_p$, let $\Psi_{\rm net}$ be the ChernFormer trained on that state.  Their normalized overlap is
\begin{equation}
 \mathcal O_p=
 \frac{|\langle\Psi_p|\Psi_{\rm net}\rangle|}
 {\sqrt{\langle\Psi_p|\Psi_p\rangle
 \langle\Psi_{\rm net}|\Psi_{\rm net}\rangle}}.
 \label{eq:overlap}
\end{equation}
The brackets denote integration over all particle coordinates.  The denominator normalizes the states, and the absolute value removes the physically irrelevant overall phase.  Thus $0\leq\mathcal O_p\leq1$, with unity meaning exact agreement.  The overlap is demanding because it compares the complete complex wave function.

The green starred curve in Fig.~\ref{fig:overlap} is the squared-PsiFormer ansatz for $p=2$; the other curves use ChernFormer. Squaring doubles small phase and log-amplitude errors at leading order, whereas the Chern--Simons layer preserves the backend overlap.  ChernFormer is therefore more accurate here and, because it also contains odd-$p$ sectors, strictly broader.

The overlap in Fig.~\ref{fig:overlap} decreases smoothly as $N$ or $p$ grows, as expected when the configuration space expands and the state acquires additional collective winding.  
Overlap alone cannot prove the winding.  A narrow, low-probability region may contain a zero and a phase slip while contributing little to Eq.~\eqref{eq:overlap}.  We therefore inspect amplitude and phase slices of the learned state.

\subsection{Seeing local and collective vortices}

The full wave function depends on $2N$ real coordinates and cannot be drawn directly.  We hold $N-1$ particles fixed and scan the remaining coordinate, chosen as $z=z_1$.  Bosonic symmetry makes this choice immaterial.  Up to a factor that depends only on the fixed particles,
\begin{equation}
 \begin{aligned}
 \Psi_p(z\mid z_2,\ldots,z_N)\propto{}&
 \left(z+\sum_{j=2}^{N}z_j\right)^p
 \prod_{j=2}^{N}(z-z_j)^2\\[-2pt]
 &\times\exp\left[-\frac{|z|^2}{4\llB^2}\right].
 \end{aligned}
 \label{eq:conditionalslice}
\end{equation}
The vertical bar means that $z_2,\ldots,z_N$ are fixed.  Each factor $(z-z_j)^2$ gives a quadratic zero and two phase turns around particle $j$.  The first factor adds a collective zero at $z=-\sum_{j=2}^{N}z_j$.  Its position depends on all fixed particles, and its multiplicity $p$ gives $p$ additional phase turns.

As shown in Fig.~\ref{fig:tomography}, for $p=0$, ChernFormer reproduces the droplet, the pair-correlation holes, and the two phase turns around each fixed particle.  The $N=20$ edge-state rows provide the larger-system test requested here.  For $p=1$, the exact and learned maps preserve the local vortices and add one collective node with one phase circulation.  For $p=2$, the deeper collective depletion is accompanied by two additional turns.  These $N=20$ results show directly that the network keeps the Laughlin liquid intact while learning the new global phase of the first two edge descendants.

\section{Discussion and outlook}
\label{sec:discussion}

In this work, we ask a basic question about neural quantum states: can one impose exact bosonic exchange symmetry without deciding in advance whether the state is a condensate, a crystal, or a chiral topological liquid?  ChernFormer provides an affirmative answer to this question. Technically, it separates two tasks.  A fixed Chern--Simons phase changes fermionic exchange into bosonic exchange, while a complex Fermionic Backbone learns the physical amplitude and the remaining many-body phase.

Away from particle collisions, multiplication by the Chern--Simons phase is unitary.  In practical terms, it preserves inner products.  The bosonic training error, normalized overlap, and fidelity are therefore exactly the same as those of the inverse-transmuted fermionic target.  The fixed phase layer does not create expressive power, but it also does not remove it.  With an unrestricted complex fermionic backend, the limiting ChernFormer family can approximate any normalizable bosonic wave function on the plane or disk.  

The only built-in local constraint is the contact zero.  This constraint is physically well matched to hard-core bosons and to Laughlin-like liquids, whose amplitudes already vanish when particles meet.  For the bosonic Laughlin pair factor, inverse transmutation produces the regular antisymmetric factor in Eq.~\eqref{eq:laughlininversepair}.  A weakly interacting condensate instead has a finite contact amplitude.  It cannot be reproduced pointwise by one smooth finite ChernFormer, but it can be approached by a sequence with an increasingly narrow pair hole.  At fixed particle number, the integrated wave function and the one-body density matrix then converge to those of the condensate, and the condensate fraction approaches one.  In two dimensions, the kinetic cost of the logarithmic healing profile can also vanish for smooth, nonsingular Hamiltonians.  The essential result is therefore simple: the short-distance contact rule does not choose the long-distance order.

This contact--order separation places conventional and topological bosonic states in one variational family.  Condensation, fragmentation, density order, supersolidity, persistent currents, and chiral topological order are all allowed in the arbitrary-capacity limit.  
The Kalmeyer--Laughlin edge tower provides a particularly sharp test because it changes a global phase quantum number without changing the local Laughlin correlations.  ChernFormer leaves the center-of-mass factor $Z^p$ in the trainable many-body amplitude and can therefore carry every integer winding $p$.  An equivariant product made from $N$ identical marked-particle factors is more restrictive and generally cannot detect these edge states.

The squared-Fermionic-Backbone ansatz is bosonic but requires an antisymmetric square root, so the Kalmeyer--Laughlin center-of-mass tower is restricted to even $p$.  It also gives the lower $p=2$ overlap in Fig.~\ref{fig:overlap}.  ChernFormer is therefore the broader and more accurate architecture in this benchmark.

Importantly, multiplication by the Chern–Simons phase maps any complex antisymmetric neural wave-function family isometrically into a bosonic family. PsiFormer is the implementation used here; other generalized-determinant backbones, including QERNEL~\cite{NazaryanFu2026}, can be substituted without changing the symmetry or $L^2$-representability arguments.

The broader message is architectural.  Exact symmetry should enforce only what physics requires and should leave collective order free to emerge.  ChernFormer does this by combining an analytic statistics map with a flexible complex many-body backend.  The result is one bosonic neural wave-function language that is locally suited to hard-core and Laughlin correlations, can approach a weakly interacting condensate to arbitrarily high integrated accuracy, and can represent the low-lying chiral edge sectors excluded by an identical-factor equivariant product.

\begin{acknowledgments}
T.A.S. conducted part of the work at the Max-Planck-Institut f\"ur Physik komplexer Systeme (MPI PKS), whose support he gratefully acknowledges. This work has been supported by the Armenian Higher Education and Science Committee under the ARPI Remote Laboratory program Grant No.~24RL-1C024.
\end{acknowledgments}


\appendix

\section{Exact function-space map}
\label{app:map}

Let $D$ be the planar region available to each particle.  The collision-free configuration space is
\begin{align}
 \Omega_N(D)&=\{\RR\in D^N:z_i\neq z_j\ \text{for}\ i\neq j\},
 \nonumber\\[-2pt]
 \Omega_N&\equiv\Omega_N(D).
 \label{eq:omega}
\end{align}
Let $\cH_A$ and $\cH_S$ denote, respectively, the antisymmetric and symmetric square-integrable wave functions on this space.  The Vandermonde factor
\begin{equation}
 \Delta(\RR)=\prod_{i<j}(z_i-z_j)
\end{equation}
obeys
\begin{equation}
 \Delta(P\RR)=\sgn(P)\Delta(\RR),
 \qquad
 \chi_{\CS}(P\RR)=\sgn(P)\chi_{\CS}(\RR),
 \label{eq:chi_exchange}
\end{equation}
where $P$ is a particle permutation and $\sgn(P)=+1$ for an even permutation and $-1$ for an odd one.

Define multiplication by the Chern--Simons phase as
\begin{align}
 (U_{\CS}\Psi_F)(\RR)&=\chi_{\CS}(\RR)\Psi_F(\RR),
 \nonumber\\[-2pt]
 (U_{\CS}^{-1}\Psi_B)(\RR)&=\chi_{\CS}^{*}(\RR)\Psi_B(\RR).
 \label{eq:ucs}
\end{align}
For $\Phi,\Psi\in\cH_A$,
\begin{equation}
 \langle U_{\CS}\Phi|U_{\CS}\Psi\rangle
 =\int_{\Omega_N}\Phi^*|\chi_{\CS}|^2\Psi
 =\langle\Phi|\Psi\rangle.
\end{equation}
Thus $U_{\CS}$ is a unitary bijection from $\cH_A$ to $\cH_S$.

For a selected PsiFormer family $\cF_{\PF}\subset\cH_A$, define its ChernFormer image by
\begin{equation}
 \cB_{\CF}(\cF_{\PF})=
 \{\chi_{\CS}\Psi_F:\Psi_F\in\cF_{\PF}\}.
 \label{eq:exactclass}
\end{equation}
For a target $\Psi_B\in\cH_S$,
\begin{align}
 \inf_{\Psi_F\in\cF_{\PF}}
 \|\Psi_B-\chi_{\CS}\Psi_F\|_2
 &=\inf_{\Psi_F\in\cF_{\PF}}
 \|\chi_{\CS}^{*}\Psi_B-\Psi_F\|_2.
 \label{eq:distanceappendix}
\end{align}
The same identity holds for normalized overlaps and fidelities.  Since a unitary map is a homeomorphism, it also transports closures:
\begin{equation}
 \overline{\cB_{\CF}(\cF_{\PF})}
 =U_{\CS}\overline{\cF_{\PF}}.
\end{equation}

\section{Constructive generalized determinant}
\label{app:determinant}

Let $K\subset\Omega_N$ be compact and permutation invariant, and let $\Psi_B$ be continuous and symmetric on $K$. The inverse target $\Psi_F=\chi_{\CS}^{*}\Psi_B$ is continuous and antisymmetric. Since $\Delta$ has no zeros on $K$, the quotient $S=\Psi_F/\Delta$ is continuous and symmetric.

Consider the generalized orbital matrix
\begin{equation}
 \Phi(\RR)=
 \begin{pmatrix}
 S(\RR)&S(\RR)&\cdots&S(\RR)\\
 z_1&z_2&\cdots&z_N\\
 z_1^2&z_2^2&\cdots&z_N^2\\
 \vdots&\vdots&\ddots&\vdots\\
 z_1^{N-1}&z_2^{N-1}&\cdots&z_N^{N-1}
 \end{pmatrix}.
 \label{eq:detmatrix}
\end{equation}
Factoring $S$ from the first row gives
\begin{equation}
 \det\Phi=C_N S(\RR)\Delta(\RR)=C_N\Psi_F(\RR),
 \label{eq:detidentity}
\end{equation}
where $C_N=\pm1$ depends on the Vandermonde convention. A permutation-equivariant transformer can approximate the required symmetric scalar and particle-wise monomials. This yields uniform approximation on $K$.

For $L^2$ density on an unbounded domain, first truncate the target to a compact set, then remove a sufficiently small neighborhood of the collision manifold, and finally approximate uniformly on the remaining compact collision-free set. The discarded tail and collision tube can have arbitrarily small $L^2$ norm. A flexible envelope restores the desired decay.

\section{Condensate closure in norm, one-body density matrix, and energy}
\label{app:condensate}

Assume a bounded planar domain and fixed $N$. Let $\Phi_0=\prod_i\phi(\rr_i)$ with bounded $\phi\in H^1(D)$. The regulator used in the main text is
\begin{equation}
 F_{\delta,\epsilon}(\RR)=\prod_{i<j}f_{\delta,\epsilon}(r_{ij}),
 \qquad r_{ij}=|\rr_i-\rr_j|,
\end{equation}
with
\begin{equation}
 f_{\delta,\epsilon}(r)=
 \begin{cases}
 0, & r\leq\delta,\\[2pt]
 \dfrac{\log(r/\delta)}{\log(\epsilon/\delta)}, & \delta<r<\epsilon,\\[7pt]
 1, & r\geq\epsilon.
 \end{cases}
 \label{eq:logholeappendix}
\end{equation}
For each fixed $0<\delta<\epsilon$, the profile in Eq.~\eqref{eq:logholeappendix} belongs to $H^1$.  For every $\eta>0$, its two corners can be rounded inside arbitrarily thin annuli to obtain a $C^\infty$ profile $\widetilde f_{\delta,\epsilon}$ with $0\leq\widetilde f_{\delta,\epsilon}\leq1$ and $\|\widetilde f_{\delta,\epsilon}-f_{\delta,\epsilon}\|_{H^1}<\eta$.  Hence all norm and energy bounds below are stable under smoothing.
The one-body density matrix is
\begin{equation}
 \gamma^{(1)}(\rr,\rr')=
 N\int\dd\rr_2\cdots\dd\rr_N\,
 \Psi(\rr,\rr_2,\ldots)\Psi^*(\rr',\rr_2,\ldots),
\end{equation}
with $\Tr\gamma^{(1)}=N$. The set
\begin{equation}
 A_\epsilon=\{\RR:\min_{i<j}r_{ij}<\epsilon\}
\end{equation}
has probability under $|\Phi_0|^2$ bounded by $C N(N-1)\epsilon^2$ for small $\epsilon$. Since $F_{\delta,\epsilon}=1$ outside $A_\epsilon$ and $0\leq F_{\delta,\epsilon}\leq1$,
\begin{equation}
 \|(1-F_{\delta,\epsilon})\Phi_0\|_2^2
 \leq \int_{A_\epsilon}|\Phi_0|^2
 \leq C N^2\epsilon^2.
\end{equation}
The normalization constant in Eq.~\eqref{eq:condensatesequence} tends to one, proving Eq.~\eqref{eq:l2condensate}.

For normalized pure states, the trace distance of the projectors is bounded by the Hilbert-space distance. Taking a partial trace cannot increase trace distance. With the convention $\Tr\gamma^{(1)}=N$,
\begin{equation}
 \left\|\frac{\gamma^{(1)}_{\Phi}}{N}
 -\frac{\gamma^{(1)}_{\Psi}}{N}\right\|_1
 \leq 2\|\Phi-\ee^{\ii\alpha}\Psi\|_2
\end{equation}
for a suitable global phase $\alpha$. The eigenvalue vectors of the normalized one-body density matrices therefore converge in $\ell^1$. Combining this bound with Eq.~\eqref{eq:l2condensate} gives, for bounded $\phi$,
\begin{equation}
 1-\frac{n_0}{N}\leq C_\phi N\epsilon
 \label{eq:appendixcondensateaccuracy}
\end{equation}
for sufficiently small $\epsilon$. This proves convergence of the condensate fraction and of fragmented macroscopic occupations, and quantifies the statement that the condensate can be represented with arbitrarily high accuracy at fixed $N$.

For the kinetic term, write $F=\prod_{i<j}f_{ij}$. The difference $\nabla(F\Phi_0)-\nabla\Phi_0$ contains $(F-1)\nabla\Phi_0$, which vanishes in norm because its support shrinks, and $\Phi_0\nabla F$. For fixed $N$, the latter is bounded by a finite sum of pair contributions. The normal-plane integral for each pair is Eq.~\eqref{eq:logenergy}. Choosing $\epsilon_n\to0$ and $\log(\epsilon_n/\delta_n)\to\infty$ gives
\begin{equation}
 \|\Phi_{\delta_n,\epsilon_n}-\Phi_0\|_{H^1}\to0.
 \label{eq:h1conv}
\end{equation}
For bounded local potentials, or smooth confining potentials with controlled tails, the energy converges. Singular zero-range interactions require separate treatment because they probe the contact set directly.

The corresponding inverse-transmuted fermion is continuous because $F$ vanishes in a neighborhood of every collision. A derivative-level universal approximation by complex PsiFormer is needed to transfer Eq.~\eqref{eq:h1conv} from the analytic sequence to an implemented network. The unconditional architecture result is the $L^2$ and one-body-density-matrix closure; the energy statement adds this Sobolev-approximation requirement.

\section{Kalmeyer--Laughlin identities and winding diagnostics}
\label{app:KL}

Introduce relative coordinates $\xi_i=z_i-Z/N$. Pair differences depend only on $\xi_i-\xi_j$, and
\begin{equation}
 \sum_i|z_i|^2=\sum_i|\xi_i|^2+\frac{|Z|^2}{N}.
\end{equation}
The edge state factorizes as
\begin{align}
 \Psi_p
 &=Z^p\exp\left[-\frac{|Z|^2}{4N\llB^2}\right]\nonumber\\
 &\quad\times
 \left[\prod_{i<j}(\xi_i-\xi_j)^2
 \exp\left(-\sum_i\frac{|\xi_i|^2}{4\llB^2}\right)\right].
 \label{eq:KLfactorization}
\end{align}
All $p$ dependence is in the center-of-mass lowest-Landau-level factor. The internal topological liquid is unchanged.

For a nonzero wave function on a closed loop $\gamma$, define
\begin{equation}
 w_\gamma[\Psi]=\frac{1}{2\pi}\Delta_\gamma\arg\Psi.
\end{equation}
Because products add phase windings,
\begin{equation}
 w_\gamma[\Psi_{\CF}]
 =w_\gamma[\chi_{\CS}]+w_\gamma[\Psi_{\PF}].
\end{equation}
A local pair loop gives one unit from $\chi_{\CS}$. The rigid center-of-mass loop in Eq.~\eqref{eq:CMloop} keeps every $z_i-z_j$ fixed, so $w_{\cm}[\chi_{\CS}]=0$ and Eq.~\eqref{eq:CMwinding} follows.

This differs from the equivariant product in Eq.~\eqref{eq:EPmain}. On the loop in Eq.~\eqref{eq:CMloop}, define
\begin{equation}
 H_i(t)=h\bigl(z_i^{(0)}+a\ee^{\ii t};\{z_j^{(0)}+a\ee^{\ii t}\}_{j\neq i}\bigr).
\end{equation}
If the product is nonzero on the loop, each $H_i$ is a nonzero closed path in the complex plane and has an integer winding
\begin{equation}
 m_i=\frac{1}{2\pi}\Delta\arg H_i(t).
\end{equation}
The product winding is $\sum_i m_i$. The marked configuration with particle $i$ distinguished can be continuously deformed into the one with particle $j$ distinguished while the unmarked coordinates remain an unordered set. If this homotopy avoids both collisions and zeros of $h$, the winding cannot change, so $m_i=m$ for every $i$. Hence
$ w_{\cm}[\Psi_{\rm EP}]=\sum_{i=1}^{N}m_i=Nm.$
This establishes Eq.~\eqref{eq:EPwindingmain}. ChernFormer has no packet structure because the full collective winding can reside in one antisymmetric many-body amplitude. The obstruction concerns exact continuous non-nodal representation on the diagnostic region. It does not by itself give a finite upper bound on an $L^2$ overlap, because a competing ansatz can change winding by placing a zero on a set of small measure.

\section{Training and Architectural Details}
\label{app:TrainingDetails}
\begin{table}[H]
\centering
\caption{Architecture and training hyperparameters}
\label{tab:hyperparameters}

\begin{tabular}{ll}
\toprule
\textbf{Hyperparameter} & \textbf{Value} \\
\midrule

Network Type           & Psiformer \\
Number of layers       & 4 \\
Number of heads        & 4 \\
Attention dimension    & 32 \\
MLP dimension          & 128 \\
Optimizer              & KFAC \\
KFAC norm constraint   & $1 \times 10^{-3}$ \\
KFAC damping           & $1 \times 10^{-4}$ \\
Learning rate          & $1 \times 10^{-3}$ \\
Batch size             & 4096 \\
Delay                  & $1.0 \times 10^{5}$ \\
Decay                  & 1 \\
Rescale input          & False \\
Layer norm             & True \\
Precision              & FP32 \\
MCMC steps btw iterations & 10 \\
Number of determinants & 4 \\
Jastrow factor         & None \\

\bottomrule
\end{tabular}

\end{table}
To train the neural-network wave function, we use a loss function composed of two complementary terms. The first term, $L_{\rho}$, measures the difference between the probability densities of the trial wave function $\psi_{\theta}$ and the reference wave function $\psi_{\mathrm{ref}}$. It is based on the logarithm of the ratio of the wave-function amplitudes, which provides a robust measure of the density mismatch while remaining sensitive even in regions where the wave-function amplitude is small,
\begin{equation}
L_{\rho}
=
\frac{1}{\mathcal{N}}
\int d\mathbf{R}\,
|\psi_{\theta}(\mathbf{R})|^2
\left(
\ln
\left|
\frac{\psi_{\theta}(\mathbf{R})}
{\psi_{\mathrm{ref}}(\mathbf{R})}
\right|^2
\right)^2
\label{eq:Lrho}
\end{equation}
where $\mathcal{N}$ is the normalization factor.

The second term, $L_{j}$, compares the phase gradients of the trial and reference wave functions,
\begin{equation}
L_{j}
=
\frac{1}{\mathcal{N}}
\int d\mathbf{R}\,
|\psi_{\theta}(\mathbf{R})|^2
\sum_{\ell}
\left|
\nabla_{\ell}\varphi_{\theta}(\mathbf{R})
-
\nabla_{\ell}\varphi_{\mathrm{ref}}(\mathbf{R})
\right|^2
\label{eq:Lj}
\end{equation}
where $\varphi_{\theta}$ and $\varphi_{\mathrm{ref}}$ denote the phases of the trial and reference wave functions, respectively, and $\nabla_{\ell}$ is the gradient with respect to the position of particle $\ell$. This term encourages the trial wave function to reproduce the local phase structure of the target state, which is particularly important for capturing vortices and other phase singularities.

The total loss function is then defined as
\begin{equation}
L = L_{\rho} + \alpha L_{j}
\label{eq:loss}
\end{equation}
where the parameter $\alpha$ controls the relative weight of the density and phase-gradient contributions.

\end{document}